\documentclass{article}
\usepackage{spconf,amsmath,graphicx}
\usepackage{amsmath,amsfonts,amssymb,amsthm,array}
\usepackage{amsmath}
\usepackage{algorithm}
\usepackage{caption}
\usepackage{subcaption}
\usepackage{algpseudocode}
\usepackage{bm}
\usepackage{color}

\theoremstyle{plain}
\newtheorem{thm}{Theorem}[section]

\newtheorem{defn}[thm]{Definition}

\theoremstyle{remark}

\newcommand{\Exp}{\mathbb{E}}
\newcommand{\Prob}{\mathbb{P}}
\newcommand{\R}{\mathcal{R}}

\newcommand{\cG}{\mathcal{G}} % graph G
\newcommand{\cV}{\mathcal{V}} % node set V
\newcommand{\cE}{\mathcal{E}} % edge set E
\newcommand{\cS}{\mathcal{S}} % cycle graph C

\newcommand{\bA}{\mathbf{A}}

\newcommand{\bH}{\mathbf{H}}
\newcommand{\bI}{\mathbf{I}}

\newcommand{\bR}{\mathbf{R}}
\newcommand{\bS}{\mathbf{S}}

\newcommand{\bW}{\mathbf{W}}

\newcommand{\eqdef}{:=}

\usepackage[colorinlistoftodos,bordercolor=orange,backgroundcolor=orange!20,linecolor=orange,textsize=scriptsize]{todonotes}

\title{A New Perspective on Randomized Gossip Algorithms}

\name{Nicolas Loizou \; \& \; Peter Richt\'{a}rik}% \sthanks{Thanks to XYZ agency for funding.}}
\address{\em School of Mathematics, The University of Edinburgh, United Kingdom\\}
\begin{document}
%\ninept
%
\maketitle
\begin{abstract}
In this short note we propose a new approach for the design and analysis of randomized gossip algorithms which can be used to solve the average consensus problem. We show how that Randomized Block Kaczmarz  (RBK) method---a method for solving linear systems---works as gossip algorithm when applied to a special system encoding the underlying network. The famous pairwise gossip algorithm arises as a special case. Subsequently, we reveal a hidden duality of randomized gossip algorithms, with the dual iterative process maintaining a set of numbers attached to the edges as opposed to nodes of the network. We prove that RBK obtains a superlinear speedup in the size of the block, and demonstrate this effect  through experiments.
\end{abstract}
\begin{keywords}
Average Consensus Problem, Linear Systems, Networks, Randomized Gossip Algorithms, Randomized Block Kaczmarz
\end{keywords}

\section{Introduction}

% In the era of big data and large scale systems, one of the biggest challenges in the area of signal and information processing is the development of fast algorithms that can solve the Average Consensus Problem (ACP) in the case of large scale network (networks with extremely large number of nodes and edges). 

The average consensus problem and randomized gossip algorithms for solving it appear in many applications, including distributed data fusion in sensor networks \cite{xiao2005scheme}, load balancing \cite{cybenko1989dynamic} and clock synchronization \cite{freris2012fast}. This subject was studied extensively in the last decade; for instance, the  seminal 2006 paper of Boyd et al.\ \cite{boyd2006randomized} on randomized gossip algorithms motivated  a fury of subsequent research and generated more than 1500 citations to date. For a survey of selected relevant work prior to 2010, we refer the reader to the work of  Dimakis et al.\ \cite{dimakis2010gossip}. For more recent results on randomized gossip algorithms we suggest \cite{jun2013performance, zouzias2015randomized, liu2013analysis}. See also  \cite{dimakis2008geographic, aysal2009broadcast, olshevsky2009convergence}. 

\subsection{The average consensus problem} \label{sec:ACP}

In the average consensus (AC) problem, we are given an undirected connected network $\cG=(\cV,\cE)$ with node set $\cV=\{1,2,\dots,n\}$ and edges $\cE$. Each node $i \in \cV$ ``knows'' a private value $c_i \in \R$. The goal of AC is for every node of the network to compute the average of these private values, $\bar{c}\eqdef\tfrac{1}{n}\sum_i c_i$, in a distributed fashion. That is, the exchange of information can only occur between connected nodes (neighbors). 

\subsection{Contributions} In this paper we revisit, from a fresh perspective, the AC problem. Our starting point is the recent observation of Gower and Richt\'{a}rik \cite{gower2015stochastic} that the most basic randomized gossip algorithm (``randomly pick an edge $(i,j)\in \cE$ and then replace the values stored at vertices $i$ and $j$ by their average'') is an instance of the randomized Kaczmarz (RK) method for solving consistent linear systems,  applied to a specific linear system encoding the AC problem. The RK method was first analyzed in 2009 by Strohmer and Vershynin \cite{RK}, and since then, there was an explosion of activity in refining,  generalizing and extending the results  \cite{RBK, zouzias2013randomized, eldar2011acceleration, liu2016accelerated,  needell2015randomized}. In this paper, we examine the Stochastic Dual Ascent (SDA) method of Gower and Richt\'{a}rik \cite{gower2015stochastic}, which includes the RK method as a special case, in the context of AC problem. We show how the complexity result of SDA implies a bound on the $\varepsilon$-averaging time which is well-known in the literature for a more restricted class of  randomized gossip algorithms. Further, we explain how SDA uncovers a fundamental but hitherto hidden duality of randomized gossip algorithms, and give a natural interpretation thereof. We then focus on a specific subclass of SDA which is identical to the randomized block Kaczmarz method  \cite{RBK} in the primal space, and which can  be interpreted as a randomized Newton method  in the dual space. In particular, we show that the method has a certain superlinear speedup property, and explain what this property means.  Finally, we perform  experiments to justify the last claim.

% In recent years, Numerical Linear Algebra community made great progress in the developing of faster methods for solving large scale linear systems. More specifically from the work of Strohmer and Vershynin \cite{RK} and the introduction of randomization in the well known Kaczmarz method the area has witnessed an explosive growth. 

\section{SDA: Stochastic Dual Ascent}

In this section we briefly review those aspects of the work of
Gower and Richt\'{a}rik \cite{gower2015stochastic} on Stochastic Dual Ascent (SDA) which we will need in the rest of the paper. 

% , in which they proved that Randomized Kaczmarz method \cite{RK} and its generality Randomized Block Kaczmarz \cite{RBK} converge with an exponential rate even in the case of linear systems with no full rank matrix. 

% The only necessary assumption for convergence is that  data matrix $\bA \in \R^{m \times n}$ has no zero rows. More specifically they show that Kaczmarz methods can solve problem \eqref{generalproblem} where $c \in \R^n$ is a fixed vector. That is they can find the solution of the linear system that is closest with respect to a Euclidean distance, to a given vector  $c \in \R^n$.

Consider an $m\times n$ real matrix $\bA$ (assume it does not contain any zero rows) and vector $b\in \R^m$ such that the linear system $\bA x = b$ is consistent (i.e., has a solution).  Since we do not assume the solution is unique, we shall be interested in a particular solution:
\begin{equation}
\label{generalproblem}
\min_{x = (x_1,\dots, x_n) \in \R^n} \tfrac{1}{2} \|x-c\|^2 
\quad \text{subject to}  \quad \bA x = b.
\end{equation}
Above, $c=(c_1,\dots,c_n)\in \R^n$ is a given vector and $\|\cdot\|$ is the standard Euclidean norm. In words, in \eqref{generalproblem} we are seeking the solution of the system which is closest to $c$.  By $x^*$ we denote the  solution of \eqref{generalproblem}. The  {\em dual} of  problem \eqref{generalproblem} is
\begin{equation}
\label{Dual Problem}
\max_{y\in \R^m} D(y) \eqdef (b-\bA c)^\top y - \tfrac{1}{2}\|\bA^\top y\|^2.
\end{equation}
SDA is a randomized iterative algorithm for solving \eqref{Dual Problem}, performing the  iteration $y^{k+1} = y^k + \bS_k \lambda^k$, where $\bS_k$ is a matrix chosen in an i.i.d.\ fashion throughout the iterative process from an arbitrary but fixed distribution (which is a parameter of the method)  and $\lambda^k$ is a vector chosen {\em afterwards} so that $D(y^k + \bS_k \lambda)$ is maximized in $\lambda$. In general, the maximizer in $\lambda$ is not unique. In SDA, we let $\lambda^k$ to be the least-norm maximizer, which leads to the iteration
\begin{equation}
\label{alg:dual}
\boxed{ y^{k+1}= y^k - \bR_k (\bA(c+\bA^\top y^k) - b)}
\end{equation}
where  $\bR_k \eqdef \bS_k((\bS_k)^\top \bA \bA^\top  \bS_k)^\dagger (\bS_k)^\top$ (this matrix is always symmetric and positive semidefinite). With the sequence of the {\em dual iterates} $\{y^k\}$ we associate a sequence of {\em primal iterates} $\{x^k\}$ as follows:
\begin{equation}
\label{connection}
 x^k\eqdef c+\bA^\top y^k.
\end{equation}
By combining \eqref{connection} with \eqref{alg:dual}, we obtain the following algorithm:
\begin{equation}
\label{alg:primal}
\boxed{x^{k+1} = x^k - \bA^\top \bR_k (\bA x^k-b)}
\end{equation}

If $\bS_k$ is chosen randomly from the set of  unit coordinate/basis vectors in $\R^m$,  then the dual method \eqref{alg:dual} is randomized coordinate descent \cite{leventhal2010randomized,  serial}, and the corresponding primal method \eqref{alg:primal} is RK.  More generally, if $\bS_k$ is a random column submatrix of the $m \times m$ identity matrix, the dual method is the randomized Newton method \cite{qu2015sdna}, and the corresponding primal method is a block version of RK \cite{RBK}. We shall describe the more general case in more detail in Section~\ref{sec:block}.

% It can be shown that if matrix $\bS$ is chosen to be a unit coordinate vector $e_i \in \R^m$ then equation \eqref{alg:dual} is equivalent with the \textit{randomized coordinate descent method}  and when $\bS$ is a random column submatrix of the $m \times m$ Identity matrix then SDA is precisely the \textit{Randomized Newton method} that first proposed in 

The basic convergence guarantees for both the primal and the dual iterative processes are presented in the following theorem.  We set $y^0=0$ so that $x^0 = c$, which corresponds to the vector of initial private values stored at the nodes.

 \begin{thm}[Complexity of SDA \cite{gower2015stochastic}]
 \label{maintheorem}
Let $y^0=0$ and assume that the matrix $\bH \eqdef \Exp[\bR_k]$ is well defined and nonsingular.  Then the dual iterates $\{y^k\}$ of SDA defined in \eqref{alg:dual} for all $k\geq 0$ satisfy
\begin{equation}\label{eq:conv-dual}\Exp[D(y^*)-D(y^k)]\leq \rho^k (D(y^*)-D(y^0)).\end{equation}
Likewise, the corresponding primal iterates, defined in \eqref{connection} and explicitly written in \eqref{alg:primal}, for all $k\geq 0$ satisfy
\begin{equation}\label{eq:conv-primal}\Exp[\|x^k-x^*\|^2]\leq \rho^k \|x^0-x^*\|^2.\end{equation}
The convergence rate $\rho$ is given by
\begin{equation}
\label{convergencerate}
\rho \eqdef 1 - \lambda_{\min}^+(\bA^\top \bH \bA) \in (0,1),
\end{equation}  
 where  $\lambda_{\min}^+(\cdot)$ denotes the minimum nonzero eigenvalue.
 \end{thm}
 
\section{Randomized Gossip \& SDA}

We propose that randomized gossip algorithms be viewed as applications of SDA (either in the primal or dual form) to a particular problem of the form \eqref{generalproblem} (resp.\ \eqref{Dual Problem}). In particular, we let $c=(c_1,\dots,c_n)$ be the initial values stored at the nodes of $\cG$, and choose $\bA$ and $b$ so that the constraint $\bA x = b$ is equivalent to the requirement that $x_i=x_j$ (the value stored at node $i$ is equal to the value stored at node $j$) for all $(i,j)\in \cE$.

\begin{defn} We say that  $\bA x = b$ is an ``average consensus (AC) system'' when $\bA x = b$ iff $x_i = x_j$ for all $(i,j) \in \cE$.
\end{defn}

It is easy to see that $\bA x = b$ is an AC system precisely when $b=0$ and the nullspace of $\bA$ is $\{t 1_n : t\in \R\}$, where $1_n$ is the vector of all ones in $\R^n$.  Hence, $\bA$ has rank $n-1$. Moreover, it is easy to see that for any AC system, the solution of \eqref{generalproblem} necessarily is $x^* = \bar{c} \cdot 1_n$ --- this is why we singled out AC systems. In this sense, {\em any} algorithm for solving \eqref{generalproblem} will ``find'' the average $\bar{c}$. However, in order to obtain a distributed algorithm we need to make sure that only ``local'' (with respect to $\cG$) exchange of information is allowed.  

\subsection{Standard Form and Mass Preservation}

Assume that $\bA x = b$ is an AC system. Then the primal iterative process \eqref{alg:primal} can be written in the form
\begin{equation}\label{eq:W_k}x^{k+1} = \bW_k x^k, \end{equation}
where $\bW_k\eqdef \bI - \bA^\top \bR_k \bA$.
Eq \eqref{eq:W_k} is the standard form in which randomized gossip algorithms are written. What is new here is that the iteration matrix $\bW_k$ has a specific structure which guarantees convergence to $x^*$ under very weak assumption (see Theorem~\ref{maintheorem}). Note that if $y^0=0$, then $x^0=c$, i.e., the starting primal iterate is the vector of private values (as should be expected from any gossip algorithm).

The primal iterates \eqref{alg:primal} of SDA enjoy a mass preservation property (the proof  follows from \eqref{connection} in view of $\bA 1_n = 0$):

\begin{thm}[Mass preservation]
If $\bA x =b$ is an AC system, then the primal  iterates \eqref{alg:primal} for all  $k \geq 0$ satisfy:  $\tfrac{1}{n}\sum_{i=1}^{n}x_i^k=\bar{c}$.
 \end{thm}

\subsection{$\varepsilon$-Averaging Time}

Let $z^k\eqdef \|x^k - x^*\|$. The typical measure of convergence speed employed in the randomized gossip literature, called $\varepsilon$-averaging time and here denoted by $K(\varepsilon)$, represents the smallest time $k$ for which  $x^{k}$ gets within $\varepsilon z^0$ from $x^*$, with probability greater than $1-\varepsilon$, uniformly over all starting values $x^0=c$. More formally, we define
%\begin{equation}
%\label{Tave}
\[
K(\varepsilon)\eqdef \sup_{c\in \R^n} \inf  \{k\;:\; \Prob \left(z^k > \varepsilon z^0 \right)\leq\varepsilon\}.
\]
%\end{equation}
This definition differs slightly from the standard one in that we use $z^0$ instead of $\|c\|$.

Inequality \eqref{eq:conv-primal}, together with Markov inequality, can be used  to give a bound on $K(\varepsilon)$, formalized next:

\begin{thm}\label{thm:complexity_standard} Assume $\bA x=b$ is an AC system. Let $y^0=0$ and assume $\bH=\Exp[\bR_k]$ is nonsingular. Then for any $0<\varepsilon < 1$ we have
$K(\epsilon) \leq 3 \log(1/\varepsilon)/ \log(1/\rho) \leq \tfrac{3}{1-\rho}\log(1/\epsilon),$
where $\rho$ is defined in \eqref{convergencerate}. 
\end{thm}

It can be shown that under the assumptions of the above theorem, $\bA^\top \bH \bA$ only has a single zero eigenvalue, and hence $\lambda_{\min}^+ (\bA^\top \bH \bA)$ is the second smallest eigenvalue of $\bA^\top \bH \bA$. Thus, $\rho$ is the second largest eigenvalue of $\bI - \bA^\top \bH \bA = \Exp[\bW_k]$. The bound on $K(\varepsilon)$ appearing in Thm~\ref{thm:complexity_standard} is often written with $\rho$ replaced by $\lambda_2(\Exp[\bW_k])$ \cite{boyd2006randomized}.

%The standard  result for a basic gossip algorithm (in ) states that \cite{boyd2006randomized}
%\begin{equation}
%\label{bounds}
%\frac{0.5 \log(1/\varepsilon)}{\log(1/\lambda_2(\bW))}\leq K(\varepsilon)\leq \frac{3 \log(1/\varepsilon)}{\log(1/\lambda_2(\bW))},
%\end{equation}
%where $\bW \eqdef \Exp[\bW^{k}]$, and $\lambda_2(\bW)$ denotes  the second largest eigenvalue of $\bW$ 

\section{Block gossip algorithms} \label{sec:block}

In the previous section we highlighted some properties of SDA relevant to the randomized gossip literature, but without interpreting SDA as a gossip, or for that matter, distributed algorithm. In this section we remedy this by focusing on a particular  AC system and a particular random matrix $\bS_k$. By being specific, we will be able to give a natural interpretation of SDA as a gossip algorithm.

In particular, we choose $\bA$ to be the $|\cE| \times n$ matrix such that  $\bA x = 0$ directly encodes the constraints $x_i=x_j$ for $(i,j)\in \cE$. That is, row  $e=(i,j) \in \cE$ of matrix $\bA$ contains value $1$ in column $i$, value $-1$ in column $j$ (we use an arbitrary but fixed order of nodes defining each edge in order to fix $\bA$) and zeros elsewhere. 

Next, $\bS_k$ is selected in each iteration to be a random column submatrix of the $m \times m$ identity matrix corresponding to columns indexed by a random subset of edges $\cS_k \subseteq \cE$. We shall write  $\bS_k=\bI_{\cS_k}$. If $\cS_k=\{1,2\}$, for instance, then $\bS_k$ consists of the first and second column of $\bI$. For simplicity, from now on we will drop the subscript and write $\cS$ instead of $\cS_k$. This choice means that primal SDA is the {\em randomized block Kaczmarz (RBK) method}.

\subsection{Randomized Block Kaczmarz as a Gossip Algorithm}
 
In our setup, the primal iterative process \eqref{alg:primal} has the form:
\begin{equation}
\label{RBKasSDA}
x^{k+1} = x^k - \bA^ \top \bI_{\cS}(\bI_{\cS}^\top \bA\bA^\top \bI_{\cS})^{\dagger}\bI_{\cS}^\top \bA x^k.
\end{equation}
Algorithm \eqref{RBKasSDA} can be shown to be  equivalent to  the following ``sketch and project'' iteration (see  \cite{gower2015randomized} for additional equivalent viewpoints):
\begin{equation}
\label{RBKaLgorithm}
x^{k+1}=\underset{x \in \R^n}{\operatorname{argmin}} \{\|x-x^k\|^2 \;:\; \bI_{\cS}^\top \bA x=0\}
\end{equation} 
which is a (more general) variant of the RBK method of Needell \cite{RBK}. More specifically, this method works by projecting the last iterate $x^k$ onto the solution set of a row subsystem of $\bA x=0$, where the selected rows correspond to a  random subset $\cS \subseteq \cE$ of selected edges.

\begin{figure}[htb]
\begin{minipage}[b]{1.0\linewidth}
  \centering
  \centerline{\includegraphics[width=8.3cm]{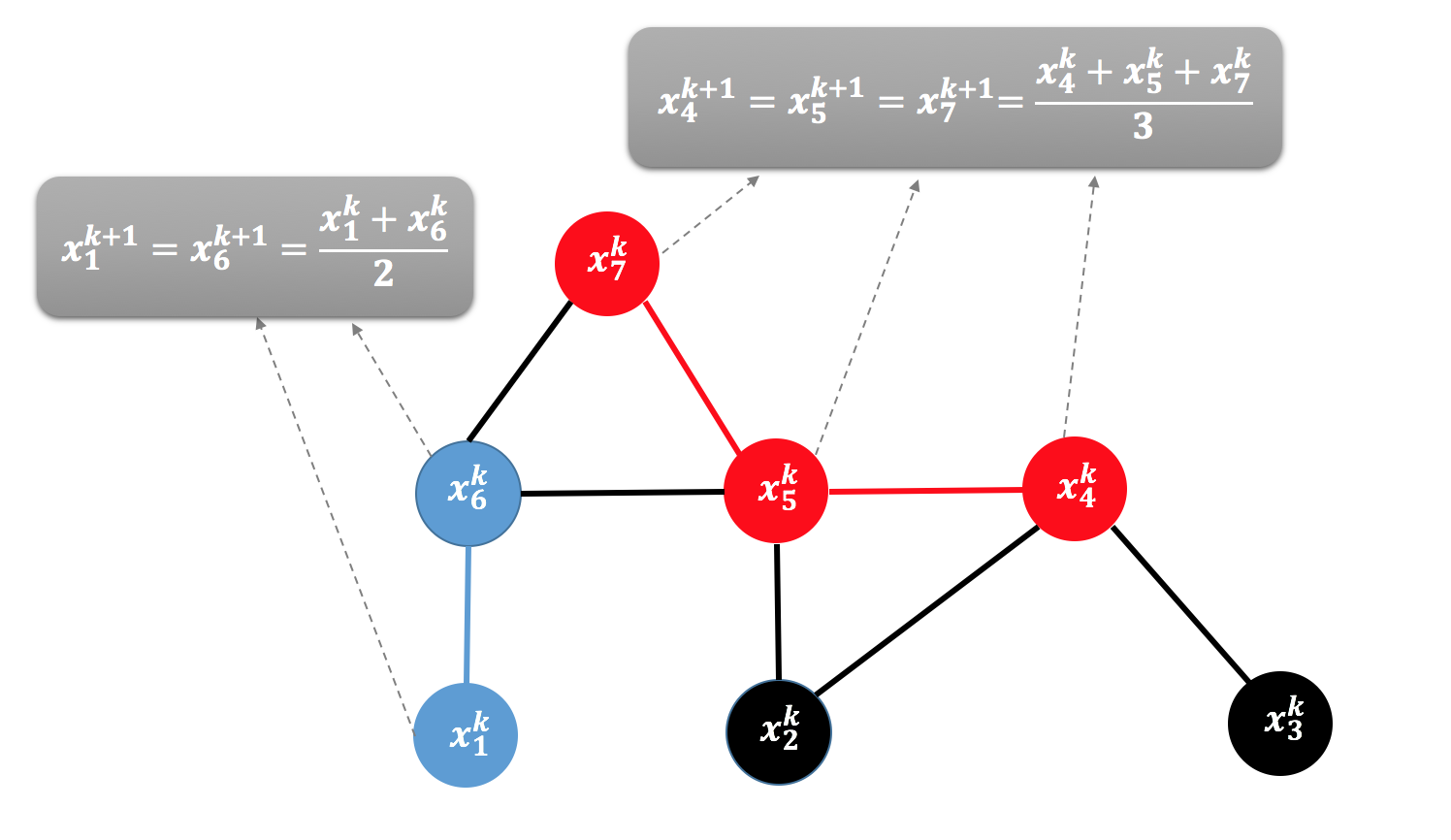}}
%  \vspace{2.0cm}
  \caption{\footnotesize Example of how the RBK method works as gossip algorithm. In the presented network 3 edges are randomly selected and a subgraph of two connected components (blue and red) is formed. Then the nodes of each connected component update their private values to their average.}
  \label{fig:RBK}
\end{minipage}
\end{figure}

While \eqref{RBKasSDA} (resp.\ \eqref{RBKaLgorithm})  may seem to be a complicated algebraic (resp.\ variational) characterization of the method, due to our choice of $\bA$ we have the following result which gives a natural interpretation of RBK as a gossip algorithm (see also Figure~\ref{fig:RBK}).

\begin{thm}[RBK as a Gossip Algorithm]
\label{TheoremRBK}
Consider the AC problem.  
Then each iteration of RBK (Algorithm~\eqref{RBKasSDA}) works as follows:
1) Select a random set of edges $\cS \subseteq \cE$, 
2) Form subgraph $\cG_k$ of $\cG$ from the selected  edges 
3) For each connected component of $\cG_k$, replace node values with their average.
\end{thm}

There is a very closed relationship between RBK and the {\em path averaging} algorithm \cite{benezit2010order}. The latter is a special case of RBK, when $\cS$ is restricted to correspond to a path of vertices.  Notice that in the special case in which $\cS$ is always a singleton, Algorithm~\eqref{RBKasSDA} reduces to the  randomized Kaczmarz method. This means that only a random edge is selected in each iteration and the nodes incident with this edge  replace their local  values with their average. This is  the pairwise gossip algorithm of Boyd \cite{boyd2006randomized}. Theorem~\ref{TheoremRBK} extends this interpretation to the case of the RBK method.

\subsection{Randomized Newton as a Dual Gossip Algorithm} 

In this subsection we bring a new insight into the randomized gossip framework by presenting how the dual iterative process that is associated to RBK method solves AC problem.
The dual iterative process \eqref{alg:dual} takes on the form:
\begin{equation}
\label{SDAsimplified}
 y^{k+1}= y^k - \bI_{\cS}(\bI_{\cS}^\top \bA \bA^\top  \bI_{\cS})^\dagger \bA(c+\bA^\top y^k).
\end{equation}
This is a randomized variant of the Newton method applied to the problem of maximizing the quadratic function $D(y)$ defined in \eqref{Dual Problem}. Indeed, as we have seen before, in each iteration we perform the update $y^{k+1} = y^k + \bI_{\cS}\lambda^k$, where $\lambda^k$ is chosen greedily so that $D(y^{k+1})$ is maximized. In doing so, we invert a random principal submatrix of the Hessian of $D$, whence the name.  

 \textit{Randomized Newton Method} (RNM) was first proposed by Qu et al.\ \cite{qu2015sdna}. RNM was first analyzed as an algorithm for minimizing \emph{smooth strongly convex functions}. In \cite{gower2015stochastic} it was also extended to the case of a \emph{smooth but weakly convex quadratics}.  This method was not previously associated with any gossip algorithm.

The most important distinction of RNM compared to  existing gossip algorithms is that it operates with values that are associated to the {\em edges} of the network. To the best of our knowledge, it the first {\em randomized dual gossip method}. In particular, instead of iterating over values  stored at  the nodes, RNM uses these values to update ``dual weights'' $y^k \in \R^m$ that correspond to the edges $\cE$ of the network. However, deterministic dual  distributed averaging algorithms were proposed before  \cite{rabbat2005generalized, ghadimi2014admm}.

% \nicolas{The reviewer number 4 mentioned 2 papers that consider primal dual approach not similar to our approach but worth somehow to mention in one line: the papers are \cite{rabbat2005generalized} in which they present a synchronous and deterministic approach and \cite{ghadimi2014admm} in which they use the same approach with ADMM for faster averaging. I think we only have space for only 1 more reference and I suggest to mention only the paper with the path averaging. If we create somehow space one of these two papers is enough to mention}

\textbf{Natural Interpretation.} In  iteration $k$, \emph{RNM} (Algorithm~\eqref{SDAsimplified}) executes the following steps:
1) Select a random set of edges $\cS_k \subseteq \cE$, 2)  Form a subgraph $\cG_k$ of $\cG$ from the selected edges, 3) The values of the edges in each connected component of $\cG_k$ are updated: their new values are a linear combination of the private values of the nodes belonging to the connected component and of the adjacent edges of their connected components. 
% (see example of Figure 2)

\textbf{Dual Variables as Advice.} The weights $y^k$ of the edges  have a natural interpretation as \textit{advice} that each selected node receives from the network in order to update its value (to one that will eventually converge to the desired average).

Consider RNM performing the $k^{th}$ iteration and let $\cV_r$ denote the set of nodes of the selected connected component that node $i$ belongs to. Then, from Theorem~ \ref{TheoremRBK} we know that $x_i^{k+1}=\sum_{i \in \cV_r} x_i^k / |\cV_r|$. Hence, by using  \eqref{connection}, we obtain the following identity:
\begin{equation}
\label{advice}
\textstyle
 (\bA^\top y^{k+1})_i=\tfrac{1}{|\cV_r|} \sum_{i \in \cV_{r}}(c_i+(\bA^\top y^{k})_i)- c_i
\end{equation}
Thus in each step $(\bA^\top y^{k+1})_i$ represents the term (advice) that must be added to the initial value $c_i$ of node $i$ in order to update its value to the average of the values of the nodes of the connected component  $i$ belongs to.

\subsection{Importance of the dual perspective}
It was shown  in  \cite{qu2015sdna} that when RNM (and as a result, RBK) is viewed as a family of methods indexed by the size $\tau = |\cS|$ (we choose $\cS$ of fixed size in the experiments),   then $\tau \to 1/(1-\rho)$, where $\rho$ is defined in \eqref{convergencerate}, decreases {\em superlinearly} fast in $\tau$. In \cite{qu2015sdna}, this was only shown for full rank $\bA$. In the next result  we extend it to AC matrices $\bA$ (which are necessarily rank-deficient).

\begin{thm}
\label{dualityTheory}
RBK enjoys superlinear speedup in $\tau$. That is, as $\tau$ increases by some factor, the iteration complexity  drops by a factor that is at least as large.
\end{thm}

\section{Numerical Experiments}

We devote this section to experimentally evaluate the performance of the proposed gossip algorithms: RBK (the primal method) and RNM (the dual method). Recall that these methods solve the same problem, and their iterates are related via a simple affine transform. Hence, all results shown apply to both RBK and RNM.

Through these experiments we demonstrate the theoretical results presented in the previous section. That is, we show that for a  connected network $\cG$, the complexity improves superlinearly in $\tau = |\cS|$, where $\cS$ is chosen as a subset of $\cE$ of size $\tau$, uniformly at random. In comparing the number of iterations for different values of $\tau$, we use the relative error  $\varepsilon=\|x^k-x^*\| / \|c-x^*\|$. We let $c_i=i$ for each node $i\in \cV$. We run RBK until the relative error becomes smaller than $0.01$.  The blue solid line in the figures denotes the actual number of iterations (after running the code)  needed in order to achieve $\varepsilon\leq 10 ^{-2}$ for different values of $\tau$. The green dotted line represents the function $f(\tau)\eqdef \frac{\ell}{\tau}$, where $\ell$ is the number of iterations of RBK with $\tau=1$ (i.e., the pairwise gossip algorithm).  The green line depicts linear speedup;  the fact that the blue line (obtained through  experiments) is below the green line points to superlinear speedup.  

\begin{figure}[!h]
\label{RingGraph}
\centering
\begin{subfigure}{.25\textwidth}
  \centering
  \includegraphics[width=1\linewidth]{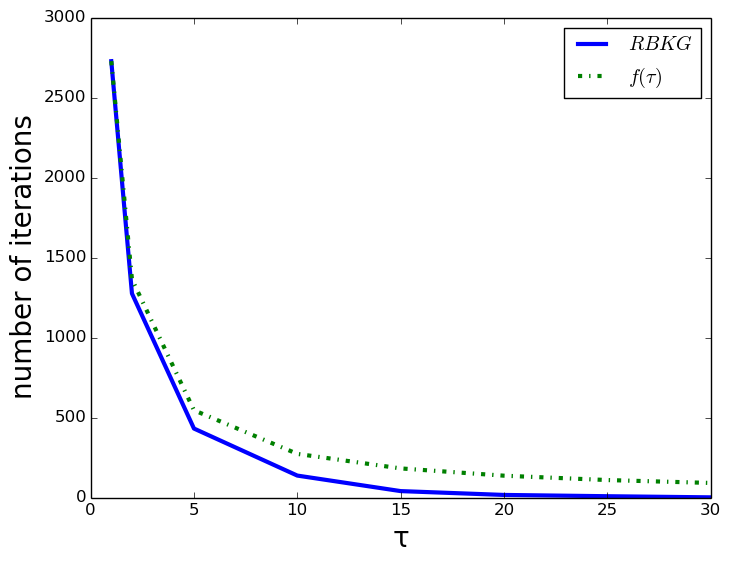}
  \caption{\footnotesize Ring graph with $n=30$}
  \label{fig:sub1}
\end{subfigure}%
\begin{subfigure}{.25\textwidth}
  \centering
  \includegraphics[width=1\linewidth]{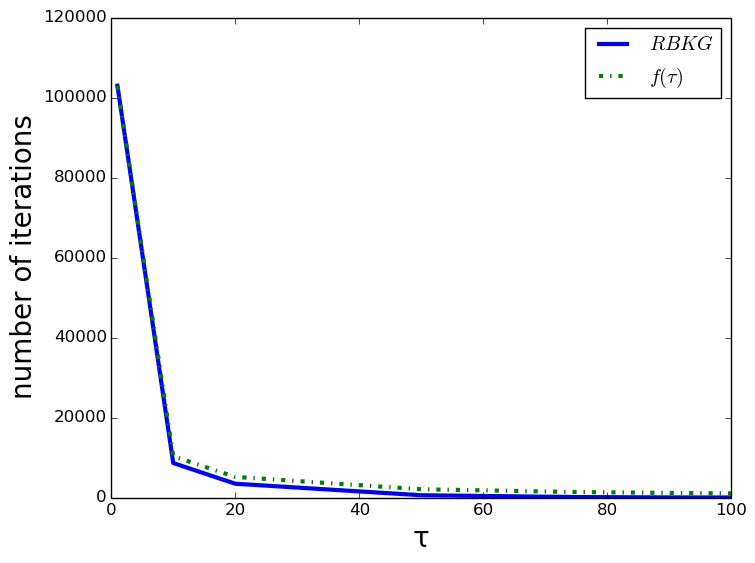}
  \caption{\footnotesize Ring graph with $n=100$}
  \label{fig:sub2}
\end{subfigure}
\caption{\footnotesize Superlinear speedup of RBK on the ring graph.}
\label{fig:test3}
\end{figure}

\begin{figure}[!h]
\label{gridGraph}
\centering
\begin{subfigure}{.23\textwidth}
  \centering
  \includegraphics[width=1\linewidth]{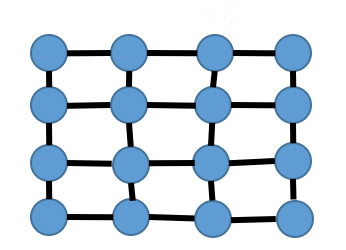}
  \caption{\footnotesize $4 \times 4$ grid graph}
  \label{fig:sub1}
\end{subfigure}%
\begin{subfigure}{.23\textwidth}
  \centering
  \includegraphics[width=1\linewidth]{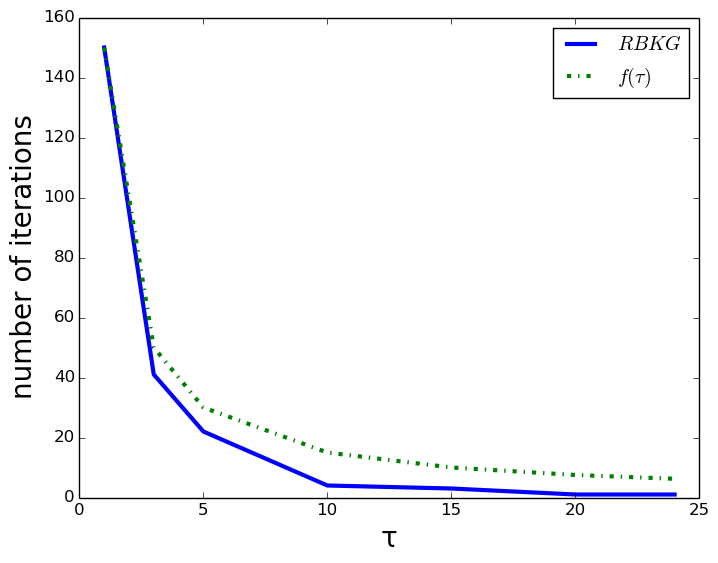}
  \caption{\footnotesize Speedup in $\tau$}
  \label{fig:sub2}
\end{subfigure}
\caption{\footnotesize Superlinear speedup of RBK on the $4 \times 4$ grid graph}
\label{fig:test4}
\end{figure}

The networks used in our experiments are the ring graph (cycle) with 30 and 100 nodes (Fig~\ref{fig:test3}) and  the $4 \times 4$ grid graph (Fig~\ref{fig:test4}). When we choose $|\cS|=m$ (i.e., we choose to update dual variables corresponding to all edges in each iteration), then $\rho=0$, and thus the method converges in one step.


% We can see that the linear speedup of RNM also holds and for the case of $\bA$ with nonzero rows that we have in our gossip framework. 

 %\section{Conclusion} In this paper we have proposed a novel approach for the design and analysis of gossip algorithms. We analyze new gossip algorithms and highlight the importance of moving to larger minibatch sizes as we can obtain at least superlinear speedup. As the sizes of networks will not stop to increase new fast methods must be develop. Our approach open up many avenues for further development and research in this area. For instance using different matrix $\bA$ can lead to faster algorithms. Furthermore, interesting results might arise if try to implement as gossip algorithms,  accelerated methods which until this point are limited to find the solution of large scale linear systems. {\color{red}depending the changes of Peter we must change the conclusion, must be as promising as possible}

% \vfill
% \pagebreak
\clearpage
\bibliographystyle{IEEEbib}
\bibliography{randgossip}

\begin{thebibliography}{10}

\bibitem{xiao2005scheme}
L.~Xiao, S.~Boyd, and S.~Lall,
\newblock ``A scheme for robust distributed sensor fusion based on average
  consensus,''
\newblock in {\em Information Processing in Sensor Networks, 2005. IPSN 2005.
  Fourth International Symposium on}. IEEE, 2005, pp. 63--70.

\bibitem{cybenko1989dynamic}
G.~Cybenko,
\newblock ``Dynamic load balancing for distributed memory multiprocessors,''
\newblock {\em J. Parallel Distrib. Comput.}, vol. 7, no. 2, pp. 279--301,
  1989.

\bibitem{freris2012fast}
N.M. Freris and A.~Zouzias,
\newblock ``Fast distributed smoothing of relative measurements,''
\newblock in {\em Decision and Control (CDC), 2012 IEEE 51st Annual Conference
  on}. IEEE, 2012, pp. 1411--1416.

\bibitem{boyd2006randomized}
S.~Boyd, A.~Ghosh, B.~Prabhakar, and D.~Shah,
\newblock ``Randomized gossip algorithms,''
\newblock {\em IEEE Trans. Inf. Theory}, vol. 14, no. SI, pp. 2508--2530, 2006.

\bibitem{dimakis2010gossip}
A.G. Dimakis, S.~Kar, J.M.F. Moura, M.G. Rabbat, and A.~Scaglione,
\newblock ``Gossip algorithms for distributed signal processing,''
\newblock {\em Proceedings of the IEEE}, vol. 98, no. 11, pp. 1847--1864, 2010.

\bibitem{jun2013performance}
J.Y. Yu and M.G. Rabbat,
\newblock ``Performance comparison of randomized gossip, broadcast gossip and
  collection tree protocol for distributed averaging,''
\newblock in {\em Computational Advances in Multi-Sensor Adaptive Processing
  (CAMSAP), 2013 IEEE 5th International Workshop on}. IEEE, 2013, pp. 93--96.

\bibitem{zouzias2015randomized}
A.~Zouzias and N.M. Freris,
\newblock ``Randomized gossip algorithms for solving {Laplacian} systems,''
\newblock in {\em Control Conference (ECC), 2015 European}. IEEE, 2015, pp.
  1920--1925.

\bibitem{liu2013analysis}
J.~Liu, B.D.O. Anderson, M.~Cao, and A.S. Morse,
\newblock ``Analysis of accelerated gossip algorithms,''
\newblock {\em Automatica}, vol. 49, no. 4, pp. 873--883, 2013.

\bibitem{dimakis2008geographic}
A.G. Dimakis, A.D. Sarwate, and M.J. Wainwright,
\newblock ``Geographic gossip: {Efficient} averaging for sensor networks,''
\newblock {\em IEEE Trans. Signal Process.}, vol. 56, no. 3, pp. 1205--1216,
  2008.

\bibitem{aysal2009broadcast}
T.C. Aysal, M.E. Yildiz, A.D. Sarwate, and A.~Scaglione,
\newblock ``Broadcast gossip algorithms for consensus,''
\newblock {\em IEEE Trans. Signal Process.}, vol. 57, no. 7, pp. 2748--2761,
  2009.

\bibitem{olshevsky2009convergence}
A.~Olshevsky and J.N. Tsitsiklis,
\newblock ``Convergence speed in distributed consensus and averaging,''
\newblock {\em SIAM J. Control Optim.}, vol. 48, no. 1, pp. 33--55, 2009.

\bibitem{gower2015stochastic}
R.M. Gower and P.~Richt{\'a}rik,
\newblock ``Stochastic dual ascent for solving linear systems,''
\newblock {\em arXiv preprint arXiv:1512.06890}, 2015.

\bibitem{RK}
T.~Strohmer and R.~Vershynin,
\newblock ``A randomized {Kaczmarz} algorithm with exponential convergence,''
\newblock {\em J. Fourier Anal. Appl.}, vol. 15, no. 2, pp. 262--278, 2009.

\bibitem{RBK}
D.~Needell and J.A. Tropp,
\newblock ``Paved with good intentions: analysis of a randomized block
  {Kaczmarz} method,''
\newblock {\em Linear Algebra Appl.}, vol. 441, pp. 199--221, 2014.

\bibitem{zouzias2013randomized}
A.~Zouzias and N.M. Freris,
\newblock ``Randomized extended {Kaczmarz} for solving least squares,''
\newblock {\em SIAM. J. Matrix Anal. \& Appl.}, vol. 34, no. 2, pp. 773--793,
  2013.

\bibitem{eldar2011acceleration}
Y.C. Eldar and D.~Needell,
\newblock ``Acceleration of randomized {Kaczmarz} method via the
  {Johnson}--{Lindenstrauss} lemma,''
\newblock {\em Numerical Algorithms}, vol. 58, no. 2, pp. 163--177, 2011.

\bibitem{liu2016accelerated}
J.~Liu and S.~Wright,
\newblock ``An accelerated randomized {Kaczmarz} algorithm,''
\newblock {\em Mathematics of Computation}, vol. 85, no. 297, pp. 153--178,
  2016.

\bibitem{needell2015randomized}
D.~Needell, R.~Zhao, and A.~Zouzias,
\newblock ``Randomized block {Kaczmarz} method with projection for solving
  least squares,''
\newblock {\em Linear Algebra Appl.}, vol. 484, pp. 322--343, 2015.

\bibitem{leventhal2010randomized}
D.~Leventhal and A.S. Lewis,
\newblock ``Randomized methods for linear constraints: convergence rates and
  conditioning,''
\newblock {\em Math. Oper. Res.}, vol. 35, no. 3, pp. 641--654, 2010.

\bibitem{serial}
P.~Richt{\'a}rik and M.~Tak{\'a}{\v{c}},
\newblock ``Iteration complexity of randomized block-coordinate descent methods
  for minimizing a composite function,''
\newblock {\em Math. Program.}, vol. 144, no. 2, pp. 1--38, 2014.

\bibitem{qu2015sdna}
Z.~Qu, P.~Richt{\'a}rik, M.~Tak{\'a}{\v{c}}, and O.~Fercoq,
\newblock ``{SDNA}: {Stochastic} dual {Newton} ascent for empirical risk
  minimization,''
\newblock {\em ICML}, 2016.

\bibitem{gower2015randomized}
R.M. Gower and P.~Richt{\'a}rik,
\newblock ``Randomized iterative methods for linear systems,''
\newblock {\em SIAM. J. Matrix Anal. \& Appl.}, vol. 36, no. 4, pp. 1660--1690,
  2015.

\bibitem{benezit2010order}
F.~B{\'e}n{\'e}zit, A.G. Dimakis, P.~Thiran, and M.~Vetterli,
\newblock ``Order-optimal consensus through randomized path averaging,''
\newblock {\em IEEE Trans. Inf. Theory}, vol. 56, no. 10, pp. 5150--5167, 2010.

\bibitem{rabbat2005generalized}
M.G. Rabbat, R.D. Nowak, and J.A. Bucklew,
\newblock ``Generalized consensus computation in networked systems with erasure
  links,''
\newblock in {\em IEEE 6th Workshop on Signal Processing Advances in Wireless
  Communications}. IEEE, 2005, pp. 1088--1092.

\bibitem{ghadimi2014admm}
E.~Ghadimi, A.~Teixeira, M.G. Rabbat, and M.~Johansson,
\newblock ``The admm algorithm for distributed averaging: Convergence rates and
  optimal parameter selection,''
\newblock in {\em 2014 48th Asilomar Conference on Signals, Systems and
  Computers}. IEEE, 2014, pp. 783--787.

\end{thebibliography}

\clearpage

\end{document}